\pdfoutput=1
\PassOptionsToPackage{dvipsnames,prologue,table}{xcolor} 
\documentclass[sigconf]{acmart}

\usepackage{algorithm}
\usepackage{algorithmic}
\usepackage{multirow}
\usepackage{enumitem}
\usepackage{caption}
\usepackage{booktabs}
\usepackage{amsmath,amsfonts}
\usepackage{textcomp}
\usepackage{subcaption}
\usepackage{footnote}
\usepackage{graphicx}
\usepackage{multirow}
\usepackage{graphicx}
\usepackage{enumitem}
\usepackage{colortbl}
\usepackage[table]{xcolor}

\allowdisplaybreaks[4]
\usepackage{hyperref}

\theoremstyle{definition}

\newcommand\M{R$^3$S}

\AtBeginDocument{%
  }

\copyrightyear{2025}
\acmYear{2025}
\setcopyright{acmlicensed}\acmConference[WWW Companion '25]{Companion Proceedings of the ACM Web Conference 2025}{April 28-May 2, 2025}{Sydney, NSW, Australia}
\acmBooktitle{Companion Proceedings of the ACM Web Conference 2025 (WWW Companion '25), April 28-May 2, 2025, Sydney, NSW, Australia}
\acmDOI{10.1145/3701716.3715547}
\acmISBN{979-8-4007-1331-6/2025/04}

\begin{document}

\title{Reward Balancing Revisited: Enhancing Offline Reinforcement Learning for Recommender Systems}

\author{Wenzheng Shu}
\authornote{Contributed equally to this research.}
\affiliation{
  \institution{University of Electronic Science and Technology of China, Chengdu, China}
  \city{}
  \country{}
}
\email{shuwenzheng926@gmail.com}

\author{Yanxiang Zeng}
\authornotemark[1]
\affiliation{
  \institution{Kuaishou Technology, Beijing, China} 
  \city{} 
  \country{}
}
\email{zengyanxiang@kuaishou.com}

\author{Yongxiang Tang}
\affiliation{
  \institution{Kuaishou Technology, Beijing, China}
  \city{}
  \country{}
}
\email{tangyongxiang@kuaishou.com}

\author{Teng Sha}
\affiliation{
  \institution{Kuaishou Technology, Beijing, China}
  \city{}
  \country{}
}
\email{shateng@kuaishou.com}

\author{Ning Luo}
\affiliation{
  \institution{Kuaishou Technology, Beijing, China}
  \city{}
  \country{}
}
\email{luoning@kuaishou.com}

\author{Yanhua Cheng}
\affiliation{
  \institution{Kuaishou Technology, Beijing, China}
  \city{}
  \country{}
}
\email{chengyanhua@kuaishou.com}

\author{Xialong Liu}
\affiliation{
  \institution{Kuaishou Technology, Beijing, China}
  \city{}
  \country{}
}
\email{zhaolei16@kuaishou.com}

\author{Fan Zhou}
\affiliation{
  \institution{University of Electronic Science and Technology of China, Chengdu, China}
  \city{}
  \country{}
}
\email{fan.zhou@uestc.edu.cn}

\author{Peng Jiang}
\authornote{Corresponding Author.}
\affiliation{
  \institution{Kuaishou Technology, Beijing, China}
  \city{}
  \country{}
}
\email{jp2006@139.com}

\renewcommand{\shortauthors}{Wenzheng Shu et al.}

\begin{abstract}
 Offline reinforcement learning (RL) has emerged as a prevalent and effective methodology for real-world recommender systems, enabling learning policies from historical data and capturing user preferences. In offline RL, reward shaping encounters significant challenges, with past efforts to incorporate prior strategies for uncertainty to improve world models or penalize underexplored state-action pairs. Despite these efforts, a critical gap remains: the simultaneous balancing of intrinsic biases in world models and the diversity of policy recommendations. To address this limitation, we present an innovative offline RL framework termed \textbf{R}eallocated \textbf{R}eward for \textbf{R}ecommender \textbf{S}ystems (\M). By integrating inherent model uncertainty to tackle the intrinsic fluctuations in reward predictions, we boost diversity for decision-making to align with a more interactive paradigm, incorporating extra penalizers with decay that deter actions leading to diminished state variety at both local and global scales. The experimental results demonstrate that \M~ improves the accuracy of world models and efficiently harmonizes the heterogeneous preferences of the users.
\end{abstract}



\maketitle

\section{Introduction}
\label{subsec:intro}
 Recommender systems play a pivotal role in helping users identify preferred items from vast catalogs, drawing substantial research interest in e-commerce. Traditionally, these systems rely on supervised learning to extract static user preferences from historical interaction logs. However, with the advent of deep learning and abundant computational resources, accurately modeling user interests has become increasingly feasible. Recently, RL has been integrated into recommender systems to capture dynamic user preferences, treating the recommendation process as an interactive challenge. This approach models user-system interactions as sequences of states, actions, and rewards, allowing RL-based methods to adapt to each interaction, rapidly respond to evolving user preferences, and enhance long-term user engagement. Despite RL's effectiveness, the dependence on expensive online interactions for training data collection remains impractical for many real-world systems~\cite{gao2023alleviating}. Thus, advancing offline RL techniques to utilize recommendation logs is essential.

\noindent \textbf{Related works.}
 Regarding offline RL, substantial research has focused on bridging the discrepancy between offline and online data. Many recent offline RL models have performed conservative value estimation by training on the observable states extracted from offline data. \textit{For example}, BCQ \cite{fujimoto2019off} employs a generative model to limit the probabilities of state-action pairs, steering clear of updates with infrequent data. CQL \cite{kumar2020conservative} imposes penalties on Q-values for state-action pairs not encountered in the data, while MOPO \cite{yu2020mopo} utilizes a pessimistic dynamics model for conservative value estimation. However, most offline RL methods aim to instill conservatism or pessimism in policies \cite{levine2020offline}, often ignoring intrinsic interactions within data distributions. Recent studies \cite{gao2023alleviating,zhang2024roler} show that utilizing world models can enhance the diversity and uncertainty of rewards. However, they only use prior empirical evidence to address the issue of reward shaping.

\noindent \textbf{Challanges.}
 The fundamental challenge in offline reinforcement learning (RL) stems from the distributional mismatch between static offline datasets and dynamically learned policies~\cite{gao2023alleviating}. This mismatch arises from two primary factors: (1) \textit{world model collapse} due to incomplete state-action coverage and \textit{excessive conservatism} in policy optimization. For the former, prior studies ~\cite{zhang2024roler} employ multiple sampling iterations from the same world model to estimate mean predicted rewards. Alternatively, probabilistic formulations of world models ~\cite{gao2023alleviating,fujimoto2019off} enable variance computation through parameterized uncertainty distributions, thus mitigating stochastic estimation errors. For the latter, a common approach is to apply a diversity entropy penalty ~\cite{gao2023alleviating} to improve interaction in strategy recommendations. Unlike previous works, we adopt diffusion-based world models to capture the inherent uncertainty of reward distributions. In addition, we extend diversity penalization by introducing attenuated interactive mechanisms. The key contributions of this work are summarized as follows:
\begin{itemize}[leftmargin=*]
    \item We propose a novel diffusion-based world model to capture the inherent uncertainty in reward prediction explicitly.
    \item To enhance the diversity of policy recommendations, we introduce a novel strategy that combines world models with agent-environment interactions.
    \item Comprehensive experiments on three real-world benchmarks show that our method is superior to the state-of-the-art methods.
\end{itemize}

\section{Preliminaries}
\subsection{Offline RL with Recommendation System}
\label{subsec:offline}
The offline RL recommendation process is fundamentally characterized by the Markov decision process (MDP) $\mathcal{M}:=<\mathcal{S}, \mathcal{A}, \hat{\mathcal{T}}, \hat{r}, \mathrm{\gamma}>$. Each $\mathrm{s} \in \mathcal{S}$ represents a user's profile, encompassing personal interests and dynamic features such as the recent interaction history. When recommending an item as an action $\mathrm{a} \in \mathcal{A}$, the system generates a scaled reward signal from user feedback, depending on the recommendation context. Since offline data lack user reactions to items uncovered, the world model $\hat{r}:=R(\mathrm{s},\mathrm{a})$ is introduced to infer users' intrinsic preferences. The transition function is defined as $\mathrm{s}':=\hat{\mathcal{T}}(\mathrm{s}, \mathrm{a}, \hat{r})$, which encodes the next state $s'$ in an autoregressive manner. Here, $\mathrm{\gamma}$ denotes the discount factor that balances the trade-off between immediate rewards and future returns. The overarching objective of the recommendation system is to learn an optimal policy that maximizes the cumulative user experience, formalized as $\arg \max_\pi \mathbb{E}_{\tau \sim \pi}[\sum_{(\mathrm{s}, \mathrm{a}) \in \tau} \mathrm{\gamma}^j \hat{r}(\mathrm{s}, \mathrm{a})]$, denoting the $j$-th cumulative reward gain under policy $\pi$ on the trajectory $\tau$ after state $\mathrm{s}$ within the MDP process $\mathcal{M}$.

\subsection{Diffusion Model for Recommendation}
\label{subsec:diffusion}
\begin{figure}[t]
    \centering
    \includegraphics[width=1.0\linewidth]{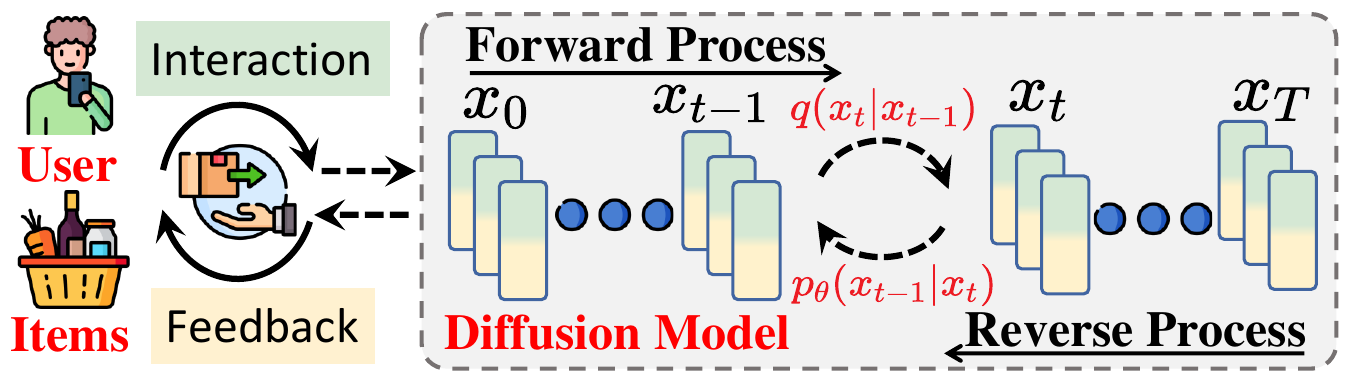}
    \caption{The overall workflow diagram of a model-based diffusion recommendation system, which takes the interaction of user and item features as input and constructs feedback (e.g., reward) through the forward and reverse processes.}
    \Description{}
    \label{fig:pics_1}
    \vspace{-4mm}
\end{figure}

The diffusion model (DM) operates through forward and reverse processes. As shown in Figure \ref{fig:pics_1}, given an initial data sample $\boldsymbol{x}_{0} \sim q(\boldsymbol{x}_{0})$, the forward process progressively adds Gaussian noise via a Markov chain over $T$ steps, generating latent variables $\boldsymbol{x}_{1:T}$. The forward transition follows $q(\boldsymbol{x}_t | \boldsymbol{x}_{t-1}) = \mathcal{N}(\boldsymbol{x}_t; \sqrt{1-\beta_t} \boldsymbol{x}_{t-1}, \beta_t \boldsymbol{I})$, where $t \in \{1, \ldots, T\}$ is the diffusion step, $\mathcal{N}$ is the Gaussian distribution, and $\beta_t \in (0,1)$ governs the noise intensity at step $t$. As $T \to \infty$, $\boldsymbol{x}_T$ tends to a standard Gaussian distribution. In the reverse process, DM reconstructs $\boldsymbol{x}_{t-1}$ from $\boldsymbol{x}_{t}$ by learning a denoising trajectory. Starting from $\boldsymbol{x}_T$, it iteratively approximates $\boldsymbol{x}_{t} \to \boldsymbol{x}_{t-1}$ through $p_\theta(\boldsymbol{x}_{t-1}|\boldsymbol{x}_{t}) = \mathcal{N}(\boldsymbol{x}_{t-1}; \mu_\theta(\boldsymbol{x}_t, t), \Sigma_\theta(\boldsymbol{x}_t, t))$, where $\mu_\theta(\boldsymbol{x}_t, t)$ and $\Sigma_\theta(\boldsymbol{x}_t, t)$ are neural network outputs parameterized by $\theta$, capturing fine-grained generative dynamics.

\section{Methodology}
\label{sec:method}
\begin{figure}[t]
    \centering
    \includegraphics[width=1.0\linewidth]{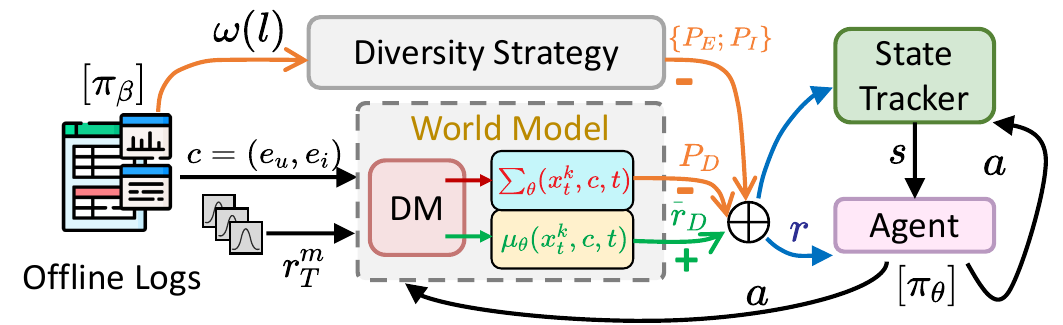}
    \caption{Outline of the ~\M~. It comprises a diffusion-based world model along with an A2C paradigm. Note that $\pi_\beta$ denotes the inherent behavior within offline data, whereas $\pi_\theta$ represents the target policy to be optimized.}
    \Description{}
    \label{fig:pics_2}
    \vspace{-4mm}
\end{figure}

 Figure \ref{fig:pics_2} depicts the overall framework, which comprises \textbf{two key stages}: (a) building a world model from offline logged data and (b) training the recommendation policy via interactions with the model. Our approach builds upon the A2C framework~\cite{mnih2016asynchronous}, extending the methodology of~\cite{zhang2024roler} to focus on reward allocation. We calibrate the world model to incorporate uncertainty in Section \ref{subsec:user_model} and apply diversity penalties to an interactive paradigm in Section \ref{subsec:diversity}.

\subsection{Inherent Diffusion Uncertainty}
\label{subsec:user_model}
 We integrate inherent uncertainty to address empirical errors and reallocate rewards, as detailed in Section \ref{subsec:intro}. Utilizing DiffRec~\cite{wang2023diffusion} as the backbone, we reconstruct the world model to generate predicted rewards $\hat{r}_D$, user $\boldsymbol{e}_u$ and item $\boldsymbol{e}_i$ embeddings as conditions, an additional uncertainty signal $P_D$.

\noindent \textbf{Foward Training.} Given the initially sparse rewards $\boldsymbol{r}_0$, we employ the forward process in Section~\ref{subsec:diffusion} to uncover intrinsic patterns in offline logged data. The combined representation $\boldsymbol{c} = \boldsymbol{e}_{u} \oplus \boldsymbol{e}_{i}$ acts as an explicit condition to guide the generative process. We optimize the model to predict $\sum_{t=2}^T \mathbb{E}_{t,\epsilon} \left[ \|\boldsymbol{\epsilon} - \boldsymbol{\epsilon}_\theta(r_t,\boldsymbol{c},\boldsymbol{t})\|_2^2 \right]$, where $\boldsymbol{r}_t$ denotes a series of corrupted rewards, $\boldsymbol{\epsilon} \sim \mathcal{N}(\boldsymbol{0}, \boldsymbol{I})$, and $\boldsymbol{\epsilon}_\theta(r_t,\boldsymbol{c}, \boldsymbol{t})$ is parameterized by the Deep Neural Network (DNN).

\noindent \textbf{Reverse Sampling.} In the reverse phase, since DM samples originate from a Gaussian distribution in a continuous space, their behavior inherently exhibits uncertainty. Instead of training additional world models as discussed in Section \ref{subsec:intro},  we perform multiple samplings to obtain a series of reward matrices. The mean reward is then updated as follows:
\begin{align}
    \label{eq:1}
    \hat{r}_D = \frac{1}{M} \sum_{m=1}^{M}\hat{W}_{\phi}[r_T^m\sim\mathcal{N}(\boldsymbol{0},\boldsymbol{I})],
\end{align}
where $\hat{W}_{\phi}[\cdot]$ denotes the total learnable parameters of our proposed world model, $r_T^m$ represents the $m$-th sample from the Gaussian distribution, and $M$ is the total number of sampling instances. Subsequently, we can directly derive $P_D$ from the predicted distributions rather than statistically evaluating the differences among distributions within additional ensembles of the world models. Considering each interaction $\boldsymbol{x}_m$, we can refine the uncertainty of reward $P_D$ directly as $\sum_{m=1}^{M} (\boldsymbol{x}_m - \overline{r}_D)^2/{M}$, which incorporates personalized perturbation information for subsequent policy learning.

\subsection{Diversity Strategy}
\label{subsec:diversity}
Leveraging the aforementioned diffusion process, we construct a more accurate world model. By simulating agent-environment interactions, we introduce additional reward penalties. Diversity integration is a critical strategy in RL, as demonstrated by \cite{gao2023alleviating}, which proposed an entropy penalty $P_E$ to mitigate the Matthew Effect:
\begin{align}
    \label{eq:2}
    P_E = -D_{\mathrm{KL}}(\pi_\beta(\cdot|\mathrm{s}) || \pi_u(\cdot|\mathrm{s})),
\end{align}
where $\pi_\beta$ corresponds to the behavior policy from offline data, $\pi_u$ represents the uniform distribution, $\boldsymbol{s}$ denotes the current state, and $D_\mathrm{KL}$ stands for KL-divergence. Building on the outstanding performance of $P_E$ as demonstrated in previous work \cite{zhang2024roler}, we integrate and advance this approach to develop a new interactive penalty, denoted as $P_I$. Additionally, we assign attenuation to the two aforementioned entropy types based on the extent of interaction.

\noindent \textbf{Interactive Penalty.} During policy iteration in interactive recommendation systems, subsequent states $\mathrm{s}'$ inevitably accumulate errors. While $P_E$ effectively enhances diversity, its dependence on a static, finite set restricts its capacity to model long-term dependencies. To address this, we propose a novel interactive penalty $P_I$, which retains the computational structure of $P_E$ but introduces a key modification: instead of calculating the $k$-order entropy from the continuous subsequence $[i-k, i-1]$ at the $i$-th iteration \((i \geq k)\), it computes entropy over $k$ randomly sampled positions within \([0, i-1]\). As $i$ increases, the diversity of samples determines the effectiveness of interactive exploration.
 
\noindent \textbf{Diversity Attenuation.}
To reconcile the potency of rewards and diversity of policy, we introduce an exponential decay function $\omega(l)=\mathrm{\alpha}(e^{-\mathrm{\xi} l}+1)$, where $l\in\{0,1,2,\cdots\}$ denotes the iteration step in trajectory $\mathrm{\tau}$. $\mathrm{\alpha}$ and $\mathrm{\xi}$ control the scale and decay rate, respectively, with values set to $\mathrm{\alpha}$ as 5e-1 and $\mathrm{\xi}$ as 1.0.

\subsection{Final Reward Paradigm}
\label{subsec:rebalance}
Integrating the insights from the previous two sections, the comprehensive reward function for overall policy learning can be expressed as follows:
\begin{align}
    \label{eq:3}
    \hat{r} = \underbrace{\hat{r}_D - \lambda_1P_D}_{UNCERTAIN} +\underbrace{\lambda_2[(1-\omega(l))P_I +\omega(l)P_E]}_{DIVERSITY}.
\end{align}
The first underbrace denotes the intrinsic diffusion uncertainty governed by predicted means and variances, essential for achieving a more robust environment. The latter combines \(P_I\) and \(P_E\) with the attenuation function \(\omega(l)\), ensuring that \(P_E\) promotes diversity in early iterations while \(P_I\) captures long-range contextual dependencies as iterations advance.

\section{Experiments}
\begin{table*}[t]
    \centering
    \setlength{\tabcolsep}{4pt}
    \renewcommand{\arraystretch}{0.5}
    \caption{Performance across three datasets is presented, with the best results highlighted in \colorbox{red!25}{red} and the second-best in \colorbox{blue!25}{blue}. Due to less overlap in item categories, $Length$ in Coat and Yahoo can easily reach their maximum interactive length.}

    \label{tab:experiments} 
    \begin{tabular}{@{}c|ccc|ccc|ccc@{}}  
        \toprule
        \textbf{Dataset}&\multicolumn{3}{c|}{\textbf{Coat}}&\multicolumn{3}{c|}{\textbf{Yahoo}}&\multicolumn{3}{c}{\textbf{KuaiRand}}\cr 
        \midrule  
        \textbf{Metric}&\multicolumn{1}{c}{$R_{tra}$}&\multicolumn{1}{c}{$R_{each}$}&\multicolumn{1}{c|}{$Length$}&\multicolumn{1}{c}{$R_{tra}$}&\multicolumn{1}{c}{$R_{each}$}&\multicolumn{1}{c|}{$Length$}
        &\multicolumn{1}{c}{$R_{tra}$}&\multicolumn{1}{c}{$R_{each}$}&\multicolumn{1}{c}{$Length$}\cr
        \midrule
        UCB&$73.671_{\pm1.811}$&$2.456_{\pm0.060}$&$30.000_{\pm0.000}$&$66.758_{\pm1.254}$&$2.225_{\pm0.042}$&$30.000_{\pm0.000}$&$1.651_{\pm0.152}$&$0.372_{\pm0.028}$&$4.431_{\pm0.212}$\cr 
        
        $\epsilon$-greedy&$72.004_{\pm1.605}$&$2.400_{\pm0.054}$&$30.000_{\pm0.000}$&$64.344_{\pm1.291}$&$2.145_{\pm0.043}$&$30.000_{\pm0.000}$&$1.711_{\pm0.126}$&$0.351_{\pm0.025}$&$4.880_{\pm0.270}$\cr
        \midrule
        SQN&$72.614_{\pm2.069}$&$2.421_{\pm0.069}$&$30.000_{\pm0.000}$&$57.727_{\pm5.751}$&$1.924_{\pm0.192}$&$30.000_{\pm0.000}$&$0.912_{\pm0.929}$&$0.182_{\pm0.058}$&$4.601_{\pm3.712}$\cr

        CRR&$67.383_{\pm1.627}$&$2.246_{\pm0.054}$&$30.000_{\pm0.000}$&$57.994_{\pm1.675}$&$1.933_{\pm0.056}$&$30.000_{\pm0.000}$&$1.481_{\pm0.124}$&$0.223_{\pm0.015}$&$6.561_{\pm0.352}$\cr
        CQL&$68.984_{\pm1.866}$&$2.230_{\pm0.062}$&$30.000_{\pm0.000}$&$62.291_{\pm3.347}$&$2.076_{\pm0.112}$&$30.000_{\pm0.000}$&$2.032_{\pm0.107}$&$0.226_{\pm0.012}$&$9.000_{\pm0.000}$\cr

        BCQ&$68.801_{\pm1.763}$&$2.293_{\pm0.059}$&$30.000_{\pm0.000}$&$61.739_{\pm1.781}$&$2.058_{\pm0.059}$&$30.000_{\pm0.000}$&$0.852_{\pm0.052}$&$0.425_{\pm0.016}$&$2.005_{\pm0.071}$\cr
        \midrule
        MBPO&$71.193_{\pm2.094}$&$2.373_{\pm0.070}$&$30.000_{\pm0.000}$&$64.550_{\pm2.157}$&$2.151_{\pm0.072}$&$30.000_{\pm0.000}$&$10.933_{\pm0.946}$&$0.431_{\pm0.021}$&$25.345_{\pm1.819}$\cr
        IPS&$73.887_{\pm1.842}$&$2.463_{\pm0.061}$&$30.000_{\pm0.000}$&$57.850_{\pm1.796}$&$1.928_{\pm0.060}$&$30.000_{\pm0.000}$&$3.629_{\pm0.676}$&$0.216_{\pm0.014}$&$16.821_{\pm3.182}$\cr
        MOPO&$71.181_{\pm2.056}$&$2.373_{\pm0.069}$&$30.000_{\pm0.000}$&$65.510_{\pm2.100}$&$2.184_{\pm0.070}$&$30.000_{\pm0.000}$&$10.934_{\pm0.963}$&$0.437_{\pm0.019}$&$25.002_{\pm1.891}$\cr
        DORL&$71.399_{\pm2.064}$&$2.380_{\pm0.069}$&$30.000_{\pm0.000}$&$66.351_{\pm2.224}$&$2.212_{\pm0.074}$&$30.000_{\pm0.000}$&$11.850_{\pm1.036}$&$0.428_{\pm0.022}$&$27.609_{\pm2.121}$\cr
        ROLeR&\cellcolor{blue!25}$76.160_{\pm2.120}$&\cellcolor{blue!25}$2.539_{\pm0.071}$&$30.000_{\pm0.000}$&\cellcolor{blue!25}$68.364_{\pm1.855}$&\cellcolor{blue!25}$2.279_{\pm0.062}$&$30.000_{\pm0.000}$&\cellcolor{blue!25}$13.455_{\pm1.509}$&\cellcolor{blue!25}$0.457_{\pm0.033}$&\cellcolor{red!25}$29.270_{\pm2.323}$\cr
        \midrule
        \textbf{\M}
        &\cellcolor{red!25}$78.224_{\pm1.643}$&\cellcolor{red!25}$2.607_{\pm0.054}$&$30.000_{\pm0.000}$&\cellcolor{red!25}$69.223_{\pm2.212}$&\cellcolor{red!25}$2.307_{\pm0.069}$&$30.000_{\pm0.000}$&\cellcolor{red!25}$14.087_{\pm1.828}$&\cellcolor{red!25}$0.488_{\pm0.045}$&\cellcolor{blue!25}$28.841_{\pm1.896}$\cr
        \bottomrule
    \end{tabular}
\end{table*}

\subsection{Experiment Settings}
\noindent \textbf{Datasets \& Evaluate Settings.} To evaluate our approach, we conduct experiments on three real-world datasets: Coat~\cite{schnabel2016recommendations} for shopping, Yahoo~\cite{marlin2009collaborative} for music, and KuaiRand~\cite{gao2022kuairand} for short videos. Details on these datasets are provided in~\cite{yu2024easyrl4rec}. We adopt the configurations from \cite{wang2023diffusion} for initial world model construction and from \cite{zhang2024roler} for the offline RL setup, ensuring a robust evaluation framework.

\noindent \textbf{Baselines \& Metrics.} We evaluate the performance of our proposed ~\M~ framework by comparing it with \textbf{eleven} RL-based methods, including \textbf{two} vanilla bandit-based approaches UCB~\cite{lai1985asymptotically} and $\epsilon$-greedy, \textbf{four} model-free methods SQN~\cite{xin2020self}, CRR~\cite{wang2020critic}, CQL~\cite{kumar2020conservative} and BCQ~\cite{fujimoto2019off}, \textbf{five} model-based methods MBPO~\cite{janner2019trust}, IPS~\cite{swaminathan2015counterfactual}, MOPO~\cite{yu2020mopo}, DORL~\cite{gao2023alleviating} and ROLeR~\cite{zhang2024roler}. We employ $R_{tra}$ (cumulative reward), $R_{each}$ (reward per step), and $Length$ (interactive distance) as our metrics, with additional details available in \cite{zhang2024roler}.

\subsection{Performance Comparison}
We evaluate the performance of various methods in different scenarios by adjusting the total expansion budget and assessing the results. The comparative results are presented in Table \ref{tab:experiments}, leading to several key observations:
(1) Model-based approaches exhibit notably superior performance compared to model-free counterparts on three datasets concerning trajectory length and cumulative rewards, especially in complex interactive environments such as KuaiRand; (2) In terms of $R_{tra}$, our method surpasses existing techniques across all datasets due to the benefits of a diffusion-based world model and diversity mechanisms in capturing reward uncertainty and interaction dynamics; (3) In KuaiRand, even without reaching the maximum interaction distance allowed by our constraints, our method achieves the highest $R_{tra}$ within those limits, resulting in a notable increase in $R_{each}$.

\subsection{Ablation Study}
To evaluate the significance of the proposed components in Equation \ref{eq:3}, we conduct relative experiments to compare the performance of our model on two typical datasets: Coat and Kuairand. The results are presented in Table~\ref{tab:ablation}. We first analyze how the diffusion world model and the improved interactive mechanism affect the overall architecture, followed by a detailed investigation into the interaction mechanism to validate the effectiveness of the two penalty mechanisms across different datasets.

\noindent \textbf{Uncertainty \& Diversity.}
We observe that the conclusions from ablation studies on two datasets are opposite: on the Coat dataset, incorporating Diffusion leads to better results attributed to the data's inherent sparsity. The inherent generalization capability of diffusion models over data distributions enables more accurate reward estimation in the world model. Consequently, we significantly outperform the DORL in cumulative rewards by replacing the world model without additional tactics. Yet, in the KuaiRand context, the benefits of Diffusion are reduced. For the setting without $\mathcal{U}$, the difference between it and DORL lies in whether $P_I$ is employed. The results show that, compared to DORL, combining entropy and interactive penalties yields higher cumulative rewards without significantly reducing interaction length.

\noindent \textbf{Different Penalizers.}
We analyze the relationships among the components within the interaction mechanism based on the diffusion world model. We can observe that using \( P_E \) or \( P_I \) alone leads to varying improvement across all metrics. In scenarios with small and sparse data scales, \( P_E \) shows better performance, while \( P_I \) excels in complex interactions. In addition, since \( P_E \) imposes stricter constraints to increase diversity, it is reasonable that its average interaction length is longer than that of \( P_I \).

\begin{table}[t]
    \centering
    \setlength{\tabcolsep}{4pt}
    \renewcommand{\arraystretch}{0.8}
    \caption{Performance of different elements setting on Coat and KuaiRand. Note that: (w/o) without ($\mathcal{U}$) diffusion uncertainty, and using DeepFM~\cite{guo2017deepfm} instead; ($\mathcal{D}$) diversity mechanism; ($P_E$) entropy penalizer; ($P_I$) interactive penalizer.}
    \label{tab:ablation}
    \resizebox{\linewidth}{!}{
    \begin{tabular}{l|ccc|l|ccc}
    \toprule
        \textbf{Coat}&$R_{tra}$&$R_{each}$&$Length$&\textbf{KuaiRand}&$R_{tra}$&$R_{each}$&$Length$\\ 
        \midrule
        w/o $\mathcal{U}$&72.369&2.412&30.000&w/o $\mathcal{U}$&12.136&0.446&27.207\\
        w/o $\mathcal{D}$&74.208&2.474&30.000&w/o $\mathcal{D}$&11.252&0.419&26.834 \\
        \midrule
        w/o $P_E$&\cellcolor{blue!25}75.433&\cellcolor{blue!25}2.514&30.000&w/o $P_E$ &\cellcolor{red!25}13.641&\cellcolor{red!25}0.482&\cellcolor{blue!25}28.296 \\
        w/o $P_I$&\cellcolor{red!25}76.243&\cellcolor{red!25}2.541&30.000&w/o $P_I$ &\cellcolor{blue!25}13.228&\cellcolor{blue!25}0.460&\cellcolor{red!25}28.782\\
    \bottomrule
    \end{tabular}
    }
\end{table}

\section{Conclusion}
In this paper, we propose~\M~ to tackle the inherent uncertainty within world models and the diversity challenges in interactive paradigms. Specifically, we first reformulate the existing world model within a diffusion framework to explicitly modulate the predicted value distribution and better capture the underlying uncertainties. Subsequently, we integrate a diversity mechanism with a decay-based weighting to assign significance to policy learning dynamically throughout the iterations. Future works will derive more diverse penalties directly from the world model, which will help enhance the learning process's diversity and extend beyond the limitations of offline datasets, thereby making the model more versatile and practical in real-world applications.

\balance
\bibliographystyle{ACM-Reference-Format}
\bibliography{reference}

\end{document}